\documentclass[12pt]{article}
\usepackage[centertags]{amsmath}
\usepackage{amsfonts}
\usepackage{amssymb}
\usepackage{amsthm}
\usepackage{newlfont}
\usepackage{stmaryrd}
\usepackage{mathrsfs}
\usepackage{pstricks,pst-node}
\usepackage{palatino}  
\usepackage{fancyhdr}
\usepackage{graphicx}
\usepackage{fancybox}
\usepackage{multibox}
\usepackage{geometry}
\usepackage[dvips]{epsfig}
\usepackage{eepic}
\usepackage{pst-plot}

\newcommand{\g}{\gamma}
\newcommand{\s}{\sigma}

\let\si=\sigma

  \let\La=\Lambda

\newcommand{\bbZ}{{\mathbb Z}}

\newcommand{\opunit}{\text{1}\kern-0.22em\text{l}}

\newcommand{\bde}{\begin{definition}}
\newcommand{\ede}{\end{definition}}
\newcommand{\beq}{\begin{equation}}
\newcommand{\eeq}{\end{equation}}
\newcommand{\ben}{\begin{enumerate}}
\newcommand{\een}{\end{enumerate}}
\newcommand{\ble}{\begin{lemma}}
\newcommand{\ele}{\end{lemma}}
\newcommand{\bpr}{\begin{proof}}
\newcommand{\epr}{\end{proof}}

\title{Two connections between random systems and non-Gibbsian measures}

\author{ 
{\normalsize Aernout C.~D.~van Enter}        \\[-1mm]
  {\normalsize\it Centre for Theoretical Physics}   \\[-1.5mm]
  {\normalsize\it Rijksuniversiteit Groningen}         \\[-1.5mm]
  {\normalsize\it Nijenborgh 4}                \\[-1.5mm]
  {\normalsize\it 9747 AG Groningen}           \\[-1.5mm]
  {\normalsize\it THE NETHERLANDS}             \\[-1mm]
  {\normalsize\tt aenter@phys.rug.nl}        \\[-1mm]
\\ [-1mm]
  {\normalsize Christof K\"ulske}            \\[-1mm]
  {\normalsize\it Department of Mathematics and Computer Science}   \\[-1.5mm]
  {\normalsize\it Rijksuniversiteit Groningen}         \\[-1.5mm]
  {\normalsize\it Blauwborgje 3}                \\[-1.5mm]
  {\normalsize\it 9747 AC Groningen}           \\[-1.5mm]
  {\normalsize\it THE NETHERLANDS}             \\[-1mm]
  {\normalsize\tt kuelske@math.rug.nl}        \\[-1mm]
{\protect\makebox[5in]{\quad}}}
\begin{document}
\maketitle \baselineskip=14pt \noindent {\bf Abstract.} In this
contribution we discuss the role disordered (or random) 
systems have played in the study 
of non-Gibbsian measures. This role has two main aspects, the distinction 
between which has not always been fully clear:

1) {\em From} disordered systems: Disordered systems can be used as  a tool; 
analogies with, as well as results and  methods from the study of random 
systems  can be employed to investigate non-Gibbsian properties of a variety
of measures of physical and mathematical interest.

2) {\em Of } disordered systems: Non-Gibbsianness is a  property of various  
(joint) measures describing   quenched disordered systems.

We discuss and review this distinction and a number of results related to 
these issues. Moreover, we discuss the mean-field version of the
non-Gibbsian property, and present some ideas how a Kac limit approach 
might connect the finite-range and the mean-field non-Gibbsian properties.

\newtheorem{theorem}{Theorem}          
\newtheorem{lemma}[theorem]{Lemma}              
\newtheorem{proposition}[theorem]{Proposition}
\newtheorem{corollary}[theorem]{Corollary}
\newtheorem{definition}[theorem]{Definition}
\newtheorem{conjecture}[theorem]{Conjecture}
\newtheorem{claim}[theorem]{Claim}
\newtheorem{observation}[theorem]{Observation}
\def\proof{\par\noindent{\it Proof.\ }}
\def\reff#1{(\ref{#1})}

\let\zed=\bbbz 
\let\szed=\bbbz 
\let\IR=\bbbr 
\let\R=\bbbr 
\let\sIR=\bbbr 
\let\IN=\bbbn 
\let\IC=\bbbc 

\def\nl{\medskip\par\noindent}

\def\scrb{{\cal B}}
\def\scrg{{\cal G}}
\def\scrf{{\cal F}}
\def\scrl{{\cal L}}
\def\scrr{{\cal R}}
\def\scrt{{\cal T}}
\def\pfin{{\cal S}}
\def\prob{M_{+1}}
\def\cql{C_{\rm ql}}
\def\bydef{\stackrel{\rm def}{=}}   
\def\qed{\hbox{\hskip 1cm\vrule width6pt height7pt depth1pt \hskip1pt}\bigskip}
\def\remark{\medskip\par\noindent{\bf Remark:}}
\def\remarks{\medskip\par\noindent{\bf Remarks:}}
\def\example{\medskip\par\noindent{\bf Example:}}
\def\examples{\medskip\par\noindent{\bf Examples:}}
\def\nonexamples{\medskip\par\noindent{\bf Non-examples:}}

\newenvironment{scarray}{
          \textfont0=\scriptfont0
          \scriptfont0=\scriptscriptfont0
          \textfont1=\scriptfont1
          \scriptfont1=\scriptscriptfont1
          \textfont2=\scriptfont2
          \scriptfont2=\scriptscriptfont2
          \textfont3=\scriptfont3
          \scriptfont3=\scriptscriptfont3
        \renewcommand{\arraystretch}{0.7}
        \begin{array}{c}}{\end{array}}

\def\wspec{w'_{\rm special}}
\def\mup{\widehat\mu^+}
\def\mupm{\widehat\mu^{+|-_\Lambda}}
\def\pip{\widehat\pi^+}
\def\pipm{\widehat\pi^{+|-_\La\bibitem{mi}
\newblock An ultimate frustration in classical lattice-gas models.
\newblock To appear in {\em J. Stat. Phys.}
mbda}}
\def\ind{{\rm I}}
\def\const{{\rm const}}

\bibliographystyle{plain}


\maketitle
\section{Introduction}
In various situations in physics where one tries to describe many-particle
systems, a description of the system as an equilibrium (Gibbs) state
in terms of an effective Hamiltonian or an effective temperature is desirable.
It has turned out, however, that such a description is not always 
available.  Indeed, one may run into non-Gibbsian states, where 
such a well-behaved Hamiltonian description simply does not exist. 

In a classical lattice set-up, non-Gibbsian states --
which then are probability measures on (spin-or particle) 
configuration spaces-- 
have been a subject of considerable interest for the last 15 
years or so, see e.g. 
\cite{EFS, Ken,Sch, LebMa,LebScho,EFHR, Isr, MaMoRe,Ferh} and also \cite{MPRF} 
and references therein.

The original study of \cite{EFS} was mainly motivated by 
Renormalization Group Transformation existence 
questions, where the findings are that
renormalized measures can be non-Gibbsian. This 
implies that in these situations 
a renormalized interaction does not exist. Thus the whole 
phenomenology of Renormalization Group fixed points, 
critical exponents expressible in terms of eigenvalues of some Renormalization 
Group operator in a space of Hamiltonians, 
its domains of attraction as universality classes, may become problematical
once one tries to justify it in a rigorous manner.

Non-Gibbsian measures showed also up in other cases of 
physical and/or mathematical interest, such as non-equilibrium states in 
either the transient or the steady-state regime, 
see e.g. \cite{EFHR,KuRe,LebScho, DR, LT}.
Yet further occurrences are for example in the study of Hidden Markov 
fields, and in Fortuin-Kasteleyn random-cluster models.

In the mathematical study  of these objects, at various occasions results 
from disordered systems, in particular the phase transition proof of the 
Random Field Ising Model and possible generalizations thereof
\cite{BK, Z} have been employed. This often is a question of mathematical 
convenience, and sometimes may be avoidable.

Also, the fact that the non-Gibbsian properties are 
due to a measure-zero set of ``bad'' configurations, that is 
spin-configurations where an presumed effective (renormalized) interaction 
diverges, shows a certain similarity to the fact, 
familiar from the study of disordered systems, that various 
physical properties of interest can be shown to 
hold only for almost all realizations of 
the disorder, but not for all of them, and that at the same time 
this zero-measure set is responsible for some subtle, more 
singular properties. 


This phenomenon is probably best known
from the analysis of the Griffiths singularity \cite{Gri}.  
This analogy has been explored in e.g. 
\cite{BCO1,BCO2,BCO3,BKL1, BKL2}.

\smallskip

A connection of different type was found first in \cite{EMSS} and then further 
explored in \cite{Ku1,Ku2,Ku3, EMK, KLR}. This is the observation that 
the quenched
(joint) measures of disordered systems  themselves have non-Gibbsian properties
and in fact this non-Gibbsianness can be more severe that what is usually found
in the Renormalization Group setting \cite{KLR}.    

Some of these considerations have been extended to a mean-field setting  where 
explicit computations can be made. We conjecture that in considerable 
generality mean-field results correspond well to the 
non-Gibbsian properties for lattice models in the limit of long-range 
(Kac-)interactions.  
 
In this paper we review a number of these aspects.


\section{ Notation and background}
For  general background on the theory of Gibbs measures we refer
to \cite{EFS, Geo}. We will mostly consider finite-spin
models, living on a finite-dimensional lattice $\bbZ^d$. The spins
will take the values in a single-spin space $\Omega_{0}$ and
we will use small Greek letters $\si,\eta,\ldots$ to denote
spin configurations for finite or infinite sets of sites. The
Hamiltonians in a finite volume
$\Lambda \subset \bbZ^d$ with boundary conditions $\eta$ outside
will be given by
\begin{equation}
H^{\Lambda}(\sigma, \eta) = \sum_{ X  \subset
\Lambda} \Phi_X([\sigma \eta] _{X})
\end{equation}
For fixed boundary condition 
these are functions on the configuration spaces.
$\Omega^{\Lambda} = \{-1,1\}^\La$.
Gibbs measures for an interaction $\Phi$ are probability measures
on $\Omega_{0}^{\bbZ^d}$ which have conditional probabilities of finding 
$\sigma$ inside $\Lambda$, given boundary condition $\eta$ outside 
$\Lambda$ which have 
a continuous version,  which are of the Gibbsian 
form 
\begin{equation}
{\frac{exp - H^{\Lambda}(\sigma, \eta)}{Z^{\Lambda}(\eta)}}. 
\end{equation}
This should be true for each volume  $\Lambda$ and  
each boundary condition $\eta$.
One requires the uniform summability condition on the interaction
\begin{equation}\label{UNIF}
||\Phi||= \sum_{0 \in X} ||\Phi_X||_{\infty} \lneq \infty 
\end{equation}

A necessary and  (nearly) sufficient condition for a probability 
measure $P$ to be such a Gibbs measure is that such conditional 
probabilities 
are continuous in the product topology - it is then 
said to to be quasilocal or almost Markovian. It 
can be shown that being Gibbsian is topologically exceptional in the set of all
(or all translation-ergodic) probability measures \cite{Isr}.

To a Gibbs measure can be associated at most one interaction. If a 
translation-invariant measure has relative entropy density zero 
(it is always larger or equal than zero) with respect to a given 
translation-invariant Gibbs measure, it then has the same large deviation 
properties and it is Gibbs for the same interaction (variational principle).

\smallskip
To show that a particular measure is non-Gibbsian, it is sufficient to 
find at least {\em one} point of discontinuity, that is at least {\em one} spin
configuration $\omega$, such that
\begin{equation}
sup_{\eta^{1}, \eta^{2}} |\mu(\sigma_0| \omega_{\Lambda}\eta^{1}_{\Lambda^c}) - \mu(\sigma_0| \omega_{\Lambda}\eta^{2}_{\Lambda^c})| \geq \epsilon, 
\end{equation}
uniformly in
$\Lambda$. Such an $\omega$ one calls a {\em bad} configuration.
In many, although not all, examples, these bad configurations have measure 
zero (almost Gibbs). A weaker property says that one can construct  
an effective potential which is summable except on a measure zero set 
(weak Gibbs) \cite{MaMoRe, KLR,EV}. If one considers transformed measures
on an image-spin space  (transformed can mean deterministically or 
stochastically renormalized or evolved, imperfectly observed....), one of 
the typical proofs of non-Gibbsianness is to consider the original 
measure {\em conditioned} on (or constrained by) an image-spin configuration. 
Once one  finds such an image-spin 
configuration for which a first-order phase transition occurs for the 
conditioned system, this usually will be a bad configuration, and the 
existence of such a bad configuration then implies non-Gibbsianness of the
transformed measure (see for details in particular \cite{EFS}, section 4.2).

For disordered models, where next to the spins there are disorder variables,
such that the interactions, Hamiltonians, measures, etc are themselves random 
objects (they are disorder-dependent), 
we will denote these disorder variables by $n$'s. 
Quenched systems will be described by probability measures on the 
pro\-duct space of the disorder and the spin variables. The disorder variables
are typically i.i.d. (they describe degrees of 
freedom which  are supposed to be 
frozen after a quench from  -- infinitely--  high temperature),
and the measures describing the quenched systems are 
defined such that the conditional measures on the spin configurations, 
given the disorder configuration, are Gibbs measures.  
Whether or not a configuration is bad now turns out to depend only on the 
disorder. That is to say, the  configuration consists of both spin 
and disorder 
variables, but  the goodness or badness of a 
particular configuration only depends on the disorder part and the spin part 
does not play any role. (This was proved in \cite{Ku1} 
under the assumption  that spin and disorder variables couple in a local way,  
which holds for all models of interest.) 

\smallskip

\section{Input from random systems for non-Gibbsianness}

\subsection*{ Finding bad configurations; maps random and nonrandom}
In various cases, the finding of a bad configuration for a given measure cannot be done explicitly. However, 
one can show that a random choice from some Bernoulli measure will do the job. That is, conditioning on a typical random 
realization of the renormalized or evolved spins  can 
induce a phase transition in the original system \cite{EFS,EFHR}, once 
it is conditioned. The existence of this phase transition often follows 
from \cite{BK,Z}. Then these typical realizations are all bad.

This strategy especially applies if one wants a result for a 
continuum of parameters, such as the
magnetic field of the untransformed measure, or the bias parameter in an 
asymmetric evolution. As one can continuously vary the mean of the Bernoulli 
measure one draws the realization from, one is then left with a continuous
family of quenched random models. 

As an example, consider a Gibbs measure for
the standard nearest neighbor Ising model in 
dimension $d$ at low temperature. Apply an infinite-temperature
Glauber dynamics (independent spin flips at Poissonian times) acting on it. 
Denote the spins at time $0$ by $\sigma$'s and the evolved spins at time $t$
by $\eta'$s.  


Then, fixing the evolved spin at site $i$
at time $t$, $\eta_i(t)$ to be plus or minus, induces an extra bias,  
that is an extra magnetic field in the plus or minus 
direction acting on the $\sigma$ spin at site $i$. The strength 
of these local (dynamical) fields  
decreases to zero with increasing time \cite{EFHR}.
Choosing the $\eta$-configuration to be alternating -a chessboard 
configuration- means that the study of the conditioned system means
considering an Ising system in an alternating field.

When the dynamical field strength $h(t)$ is weak enough -- 
which corresponds to the time being large 
enough--,  there is a phase transition in the conditioned system 
for any dimension at least $2$. 
Thus the chessboard configuration is in that case a bad configuration for the 
evolved measure. 

If, on the other hand, instead of a chessboard configuration 
one would choose a random realization from a symmetric Bernoulli measure, we 
obtain an Ising model in a random, instead of an alternating , magnetic field 
and one  obtains a phase transition for a measure-one set of such realizations,
but now only in dimension 3 or more \cite{BK}. In this way one obtains an 
uncountable set of bad configurations. 

As a side remark, note  that in general any finite-volume perturbation of a 
bad configuration is again a bad configuration. This also gives rise to 
an infinite class of bad configurations 
associated to every bad configuration.  This observation, however, only 
provides a countable set starting from a single configuration. 

If the original system was subjected to a weak external magnetic field, one  
can use a choice from a biased (in the opposite direction) 
Bernoulli measure, to compensate for this, and again obtain a phase transition.
This means considering an Ising model in a non-symmetrically distributed
random field, and, as announced in \cite{Z},  the results of \cite{BK}
still hold. In this situation
one can at the same time vary the external field and the mean 
of the compensating Bernoulli measure, and one obtains 
in this way non-Gibbsianness for a continuum of values of 
the original magnetic fields. 
This works as long as the dynamical field strength is neither much stronger 
nor much weaker than the initial field, thus the evolved measure is 
non-Gibbsian for a finite time interval. 

(In physical terminology, when one tries to  
heat the  system fast from cold to hot -- the limit measure to which one 
converges exponentially fast is a Gibbs measure, at infinite, thus very high,
temperature  --, 
non-Gibbsianness means that at intermediate times one has {\em no}
temperature, rather than an intermediate temperature.) 

The loss of the Gibbs property has also been proved for a 
diffusive time-evolution, starting from a low-temperature phase, 
for a model of continuous unbounded 
spins in a double-well potential \cite{KuRe}. This is a related example, 
where non-Gibbsian\-ness, however, does not rely on 
random field arguments, and is proved to appear in dimensions
$d\geq 2$. 
Indeed, in this model one can always  find a bad conditioning configuration 
$\eta$ for the 
time-evolved measure that is homogenous in space. 
This is possible because of the continuous nature of the spins. To prove 
non-Gibbsianness  it suffices to consider (essentially) translation-invariant 
models and avoid disordered models. 

Another difference with the results for the Ising model 
is that there is no recovery of the Gibbs-property for 
large times, in the case of non-zero initial magnetic field. Indeed the 
bad configurations are diverging with time to infinity, 
which is only possible for unbounded spins. 

Arguments which are mathematically very similar 
to the case of the Glauber time-evolution apply also 
for instance when one considers 
the behaviour of Gibbs measures under a decimation transformation \cite{EFS}.
In this case, the physical conclusion reads that a 
renormalized interaction does not exist. 

We remark that the {\it random} Glauber evolution map (which can also be seen as a 
single-site  random renormalization map) and the {\it deterministic} 
decimation map show very much the same kind of behaviour.


\bigskip

\subsection*{Analogies: measure-one properties and multiscale methods}

In various models it can be shown that the set of bad configurations 
has (non-Gibbsian) measure zero ("almost Gibbs"). 
Even if this it not true, one may  in given models 
construct an interaction which is defined almost surely 
("weak Gibbs"). By abstract arguments almost Gibbs essentially 
implies weak Gibbs \cite{MaMoRe}.

It has been observed  that there is
a similarity with  phenomenona one knows from disordered systems where the 
physics is descibed by the behaviour of the system 
for  a set of measure one from the disorder realizations. One is then 
interested in the typical (probability one) behaviour of the system.

In fact the similarity goes further, in the sense that certain configurations 
can locally be good at different scales. E.g. in the Griffiths singularity 
problem, a disorder configuration configuration consists of occupied clusters,
and  one finds occupied clusters at arbitrary scales, while in the 
non-Gibbsian set-up, one can find configurations where the effective 
interaction reaching the origin has ranges at various scales. This observation 
has led to similar techniques being applied to both types of problems. 
In particular the method of multiscale cluster expansions has turned out 
especially  useful in both the study of disordered systems and the study of
non-Gibbsian but weak Gibbsian measures. For some examples see e.g. 
\cite{BCO1,BCO2,BCO3, BKL1,BKL2,BuvL,FI, ENS}.


\smallskip

\section{Non-Gibbsian properties of  random systems}

\subsection*{Quenched measures and the Morita approach}

In \cite{EMSS}, it was discovered that if one considers the joint measure
of a site diluted Ising model, given by the Hamiltonian
\begin{equation}
-H(n, \sigma)= \sum_{<i,j>} n_i n_j \sigma_i \sigma_j
\end{equation}
where the $n_i$ are $0$ or $1$ with probability $p$ or $1-p$, and the 
conditional measure on the spins, given the realization of the occupation
variables is of the Gibbsian form, this joint (quenched) measure on   
 the product space of the disorder 
(occupation) variable space and the spin space, which is the limit of 
\begin{equation} \label{666}
K(n, \sigma) = P(n) \frac{exp -H(n, \sigma)}{Z(n)}
\end{equation}
is itself not a Gibbs measure as defined above. 
Informally this means that it can {\em not} be written as
\begin{equation} 
K(n, \sigma) =  \frac{exp -\bar H(n, \sigma)}{Z}
\end{equation}
in the thermodynamic limit  with a uniform summable Hamiltonian 
$\bar H(n,\sigma)$,    (which should be  
a function of  the pair $(n,\sigma)$ of 
disorder variables $n$ and spin-variables $\sigma$).


 In the language of disordered systems, the quenched measure 
cannot be written as an annealed  Gibbs  measure 
 for a proper potential  depending on the joint variables $(n,\sigma)$. 

{\bf Warning:} We use the terms ``quenched'' and ``annealed'' in our paper 
in the original sense, as describing either fast or slowly cooled systems, as
is the standard usage in the (mathematical) physics literature on
spin systems. 
Unfortunately, in some probabilistic literature 
``quenched'' is used for almost sure, and  ``annealed'' for averaged   
properties. To avoid confusion we stress that 
this is {\em not} our convention.
\medskip

A bad configuration  in this model  is for 
example an occupation configuration of two infinite occupied clusters, 
separated by an infinite empty interval of thickness one (and spin 
configuration arbitrary). To see where the nonlocality  comes from, for 
simplicity we first consider the $T=0$ case. 
Then the ground state of the Ising model on these
two semi-infinite clusters is fourfold degenerate.
Once one connects the two clusters, it is twofold degenerate.
Adding an occupied site in this interval can thus lower 
the entropy by a finite term $ln 2$ or not, depending on whether the two 
clusters have another connection (which can be arbitrarily far away)
or not. At sufficiently low temperatures a similar reasoning implies 
non-Gibbsianness of the quenched measure, as the extra free energy due to 
adding a single site can depend nonlocally on the occupation variable far away.
Note that the goodness or badness of an $(n,\sigma)$-configuration
is due to a nonlocal behaviour of the random partition function $Z(n)$, 
and only depends on the disorder variable $n$, but not on $\sigma$.

Afterwards in \cite{Ku1} criteria for the absence of the Gibbs property for 
measures on the  $(n,\sigma)$ product space of a general class of quenched 
disordered models $\mu[n]$ depending on disorder variables 
$n$ were given.  

A good example for this is the random field Ising model, given 
by the Hamiltonian 
\begin{equation} 
-H(n, \sigma)= \sum_{<i,j>}\sigma_i \sigma_j + h \sum_{i}n_i\sigma_i
\end{equation}
where the random fields $n_i$ are $1$ or $-1$ with probability $\frac{1}{2}$. 

Again we look at the large-volume limits of 
\begin{equation}\label{RFIMJM}
K(n, \sigma) = P(n) \frac{exp -H(n, \sigma)}{Z(n)}
\end{equation}
for different spin-boundary conditions. 

The class of systems allowed in the analysis of \cite{Ku1}
includes also the site-diluted Ising model for arbitrary dilutions $p$, 
(while \cite{EMSS} was restricted to the small-$p$ regime), 
bond disordered  models etc. It was shown in considerable generality
that the failure of this Gibbs property in product space occurs 
whenever there is a discontinuity in the quenched expectation $\mu[n]$ 
of the spin-observable in the Hamiltonian that is conjugate to 
the disorder variable $n$. 
In the case of the random field Ising model, this is just the expectation 
of the spin $\sigma_i$ taken w.r.t. the random-field dependent Gibbs measure 
$\mu[n](\sigma_i)$, viewed as a function  of the random fields $n$. 
In 3 dimensions a typical random field configuration 
allows both a plus and a minus state (as proved by \cite{BK}), for small 
enough $h$ and small enough temperature.   As changing
the field outside some volume $\Lambda$ to either plus or minus picks out one 
of the two possibilities, here in fact the set of bad configurations (points of
essential  discontinuity) has full measure, w.r.t. the joint measure. 
It consists of a full measure set
of fields combined with any spin configuration.  

In the physics literature,  such an 
annealed description of quenched disordered systems
in terms of an effective (so-called ``grand'') 
potential had been introduced by Morita long ago, 
and has been reinvented  and studied at various occasions, see e.g 
\cite{Mor, Kue,KM} and also \cite{Kue2, Ku3}.
As  a uniformly convergent 
grand potential does not exist in many examples, 
a controlled application of the Morita method is thus quite problematical.

What happens, however, if one gives up {\it  any} assumptions on  
the speed of convergence of the potential 
depending both on spin and disorder variables, and asks 
only for convergence, possibly arbitrarily slow, on a set of full measure?  
In \cite{Ku2} the existence of such a potential was shown 
by soft (martingale) arguments for disordered general models. 
The proof however exploits the product structure of the model, so that  
this it would not generalize (say) to renormalized measures. 
To summarize, for quenched models 
one loses in general any control when one truncates 
the Hamiltonian, but it least there is a well-defined Hamiltonian
to talk about. 

To get any bounds on the speed of convergence of the "grand potential",  
even on a restricted set of realizations of the disorder,  is difficult and 
model-dependent work is needed that can be hard. 
In the case of the random field Ising model, 
this can however be done \cite{KLR}, 
building on the renormalization group arguments of \cite{BK}.  
It shows  the  decay of the potential like a stretched exponential 
for almost any random field configuration $n$.



\subsection*{Almost versus weak Gibbs, violating the variational principle 
}

We just saw that although the quenched measures of the RFIM are weak 
Gibbsian with an almost surely rapidly decaying potential, 
they have a measure-one set of bad configurations.  
(This holds in $3$ dimensions, 
at low temperature, and small random fields.)  
The possibility of having a measure-one set of bad configurations, 
can have severe consequences: 
 A particularly surprising fact is 
that in this situation the variational principle can be violated \cite{KLR}. 
This means that the almost surely
defined potentials for the plus measure and the minus measure are not the same,
even though the relative entropy  density between these two 
translation-invariant measures is zero.  
In more physical terms,
the different phases have different (almost surely defined) grand potentials.  

This is a sharp contradiction to classical Gibbs formalism built 
around the notion of a uniformly summable potential (\ref{UNIF}). 
Here the  relative entropy density between two measures 
vanishes if and only if  they have the same conditional probabilities 
(or in other words: equivalent interaction potentials). 
The random field Ising model thus clearly shows that 
the proposed class of weak Gibbs measures is too broad for a variational 
principle to hold.  
One needs to strive for a smaller class. Partial, but not final,  results 
have been obtained in this direction: 
On the positive side, \cite{EV} 
showed the validity of the 
classical variational principle assuming concentration properties 
on certain "nice" configurations (invoking assumptions which are weaker than 
almost Gibbsianness -- where it follows from \cite{MaMoRe}--, 
in the spirit of, but stronger than, weak Gibbsianness). 



\section{Mean field}

A related analysis has turned out to be possible for mean-field models
of Curie-Weiss type.
Product states are trivially Gibbs, and non-trivial combinations 
thereof have a full set of discontinuity points, and thus are non-Gibbs
in a strong sense \cite{EL}. 
However,  the interesting approach turns out to 
replace   continuity of the conditioning in the product topology 
by continuity properties of conditional probabilities  as a 
function of {\em empirical averages} \cite{Ku3,HK,KL}. One looks at conditional probabilities in 
finite volumes and then takes the limit, rather 
than immediately considering the infinite-volume measures. 
{\em One} value of the empirical average for 
the magnetization now corresponds with a whole collection of spin 
configurations.  
The results are often of similar nature as in the short-range situation, 
but are more complete in the sense that in many cases the whole phase diagram can be treated. 
The parallel holds, 
to the extent that the "hidden" phase-transitions that are responsible 
for the discontinuities that cause non-Gibbsianness on the lattice 
occur in the same way as for the 
mean-field counterpart,  assuming e.g. large enough lattice dimensions.  Remember however that non-Gibbsianness 
is expressed in different topologies.

Properties like the full-measure set of discontinuities in the quenched 
measures in the random field Ising model reappear as large-volume asymptotics 
of finite-volume probabilities of "bad sets" \cite{Ku3,Ku4}.
 
Mean-field systems can serve as an illustration but also 
as a source of heuristics, suggesting new and sometimes unexpected 
mechanisms of non-Gibbsianness, as in the example of time-evolved measures 
\cite{KL}, where a symmetry-breaking in the set of bad configurations appears. 
Thus mean-field computations can be a fertile source,  
motivating further research.

\subsection*{Mean field - via Kac limits }

Let us discuss the relation between the notion of 
non-Gibbsianness on the lattice and in mean field in some more 
detail. 
A priori it might not be clear that there should be a close 
connection, since different topologies are involved
when looking at the continuity of various conditional 
probabilities.  
To draw a link between lattice and mean-field properties (also) on the level 
of the present discussion of non-Gibbsianness it should be 
very interesting to investigate Kac-models. 
Kac-models, going back to  \cite{KUH}, are defined in terms of
long-, but finite range interactions of the form  $J_\g(r)\equiv \g^dJ(\g r)$. 
Here $J(x)\geq 0$ is a nice function, rapidly decaying or of bounded support 
with $\int  J(x)d^d x=1$. 
A main example is the ferromagnetic Kac-Ising model of the form 
\begin{equation} \label{KacIsing}
-H_{\gamma}(\sigma)= \frac{1}{2}\sum_{i,j} J_\gamma(i-j) \sigma_i \sigma_j+h\sum_{i}\sigma_i
\end{equation}
We shall argue that in many cases 
the Gibbsian/non-Gibbsian properties 
should be compatible in a nice way 
via the Kac limit.  
More precisely, consider 
a translation-invariant lattice model with long-range (Kac-)
potential, depending 
on a number of parameters describing 
the interaction, such as e.g. temperature and field for an Ising model.
Look at a stochastic (or deterministic) 
transformation of the Gibbs measures, and identify 
the sets $\text{nG}(\g)$ of the parameter space (such as the space spanned by 
the temperature and field variables) for 
which the image measure is non-Gibbs as a lattice-measure, 
in the sense of the product topology. 
Define $\text{nG(MF)}$ as the parameter set for 
which the (corresponding) image measure   
of the (corresponding) mean-field model is 
non-Gibbs. Then, we expect the following,   
expressing Kac-compatibility of non-Gibbsianness.   

\begin{conjecture}
"Usually" 
\begin{equation} 
\text{nG}(\g) \rightarrow \text{nG(MF)}
\end{equation}
in the Kac-(long range)-limit $\g \downarrow 0$. 
\end{conjecture}

Such a statement is reminiscent of 
the stability of the phase-diagram at low temperatures 
(proved in the framework of Pirogov-Sinai theory), 
structural stability in bifurcation theory etc. 
A precise statement will depend on the transformation 
and on the precise definition of the models under consideration. 
 In particular one may need
sufficiently high lattice dimensions. 
To develop full proofs presumably will be non-trivial and at present 
 the above conjecture is more a research program then a theory.  

We will however describe in the examples below what the theorem means more 
precisely and which steps at this point are missing.

\subsubsection*{Kac-limit background} 

Let us start by reviewing what is the relation between 
translation-invariant Kac-models and mean-field models,  see also 
\cite{BoKu05}.
A general motivation for the introduction of Kac-models is the hope that by
taking first the infinite-volume limit, and then the Kac-limit 
$\gamma\downarrow 0$, 
one obtains the corresponding mean-field model. Historically this 
was motivated first by the desire to justify the Maxwell construction 
(equal-area rule) from a microscopical model. 
 The first fundamental result relating Kac and mean field and expressing 
a "Kac-compatibility" 
is the Lebowitz-Penrose theorem \cite{LP}. 
It says that the free energy of the  lattice-model 
converges to the convex (envelope of the) free energy of the corresponding  
mean-field model. 
The Lebowitz-Penrose theorem  holds in considerable generality for various
lattice spin models with short-range interactions, in any dimension. 

However, as the example of ferromagnetic Kac-model 
in 1d shows, convergence to a model with flat mean-field 
rate-function (a free energy with a phase transition) 
does not need to be accompanied 
by non-uniqueness of the infinite-volume Gibbs measure on the lattice.  
Indeed, to understand the Gibbs measures is 
a much more subtle (and dimension-dependent) question.
While uniqueness 
holds in one dimension for any finite $\g$, 
the Gibbs measure behaves in a non-trivial way and concentrates at a 
mesoscopic scale  on profiles with jumps  between values close to 
the corresponding (positive or negative) 
 mean-field magnetisation  
(\cite{COP},\cite{CaOrPi99}).  

In lattice dimensions $d\geq 2$ however, Kac-compatibility is expected to hold 
for translation-invariant ferromagnetic models also on the level of the phase 
diagram, that is, the set of Gibbs measures. 
In the special case of the standard Kac-Ising model 
(\ref{KacIsing})  this problem is indeed settled.  
The existence of ferromagnetically 
ordered low-temperature states in the Kac-model,   and moreover 
the convergence of the critical inverse temperature $\beta_c(\g)\rightarrow \beta_c(\text{MF})=1$ 
was proved independently in (\cite{BoZa1} and \cite{CMP}). Both proofs are 
based on spin-flip symmetry
and don't generalize to models without symmetry between the phases, see however
\cite{BoZa2}.  We stress that the Kac-model might behave different 
from a nearest neighbour model. An example 
of this is the new result of \cite{GoMe} on two-dimensional 
three-state Potts models.    
For a recent general description of Kac-limit results we refer to \cite{Pres}.

\subsubsection*{Decimation of standard Kac-Ising model to a sublattice}

As our first illustration we 
consider now the decimation transformation of the ferromagnetic Kac-Ising 
model to a sublattice $S$ that is kept fixed, 
independently of the Kac-parameter 
$\gamma$. Denote by $1-p$ the density of $S$ in $\bbZ^d$. 
Think of $d=2$ and the sub-lattice $S=(2\bbZ)^2$ for concreteness, so that $p=\frac{3}{4}$.
Denote by $nG(\g)$ the range of critical inverse temperatures 
for which the projected measure is non-Gibbsian. 
We will argue that 
\begin{equation}\label{KID} 
\text{nG}(\g) \rightarrow [\frac{1}{p},\infty)=\text{nG}(\text{MF})
\end{equation}
as $\g \downarrow 0$. 
To analyze Gibbsianness of the transformed Gibbs 
measures before the Kac-limit, and analyze 
badness of a configuration $\eta$, we go through the standard program for 
lattice systems. 
This leads us  to analyze the quenched model with Hamiltonian for 
the a spin-model on the "three-quarter lattice" $S^c=\bbZ^d\backslash S$ given by 
\begin{equation} \label{quenchedKI}
-\frac{1}{2}\sum_{i,j\in  S^c} J_\gamma(i-j) \sigma_i \sigma_j
-\sum_{i\in S^c}\Bigl(\sum_{j\in S}J_\gamma(i-j)\eta_j\Bigr)\sigma_i
\end{equation}
Consider a checkerboard-configuration $\eta=\eta_{\text{spec}}$. We claim 
that it is a bad configuration for low temperatures at sufficiently 
small $\g$.  Indeed, the $\eta$-dependent 
term in the Hamiltonian is neutral and translation-invariant, 
so that we can expect phase-coexistence 
for the quenched model (\ref{quenchedKI}).
Note that the site-dependent effective 
magnetic field $\sum_{j\in S}J_\gamma(i-j)\eta_j$ acting on the spin $\sigma_i$ 
becomes small uniformly in the Kac-limit, and so  
a mild modification of the contour-construction given in 
\cite{BoZa1} or \cite{CMP} should prove that there is indeed 
phase coexistence at low temperature. 

What about the critical inverse temperature  $\beta_c(\g,S^c,\eta_{\text{spec}})$?
We see that the effective 
interaction-strength on the three-quarter lattice $S^c$ is reduced 
by the knocking out of  $S$, in this way 
reducing the effective inverse temperature of the model to $p\beta$. 
Again, using (the modification of) \cite{BoZa1,CMP} up to the critical point should prove indeed 
$\beta_c(\gamma,S^c,\eta_{\text{spec}})\rightarrow \frac{1}{p}=\frac{4}{3}$.  
If we further assume that the "worst" bad configuration is indeed given by the checkerboard 
configuration, we have indeed proved the l.h.s. of (\ref{KID}).

Finally consider the corresponding mean-field set-up at zero external field. 
Here a Curie-Weiss model of size $N$ is projected 
to a subset of size  $(1-p)N$, and continuity of the single-site 
conditional probabilities of the projected measure at fixed magnetisation of the conditioning
is investigated. A computation shows
that the range of inverse temperatures 
for which non-Gibbsianness holds is indeed 
given by $[\frac{1}{p},\infty)$ (see \cite{Ku3}).
In this example, we do not need the disordered model as a tool  
for the mean-field analysis, in contrast to what happens on the 
finite-dimensional and Kac models. This is the case because conditioning
on a finite fraction of the spins being fixed does not depend on the location 
of these spins; indeed in mean-field models talking about the location of 
spins on a lattice does not make sense.

Let us contrast this with a result from \cite{KuJSP2001}. 
Here the following opposite result was proved: Start with a Kac-Ising 
model, with or without site-dependent 
magnetic fields. Take block-averages over blocks with width $l$ (sufficiently) 
smaller than the range $\frac{1}{\gamma}$ of the Kac-potential. 
Then the resulting measure is Gibbs.

\subsubsection*{Decimation of Kac models without symmetry}

Let us  consider more generally Hamiltonians of the form 
\begin{equation}
\begin{split}
&H(\s)=\frac{1}{2}\sum_{i,j}\Phi^\gamma_{i-j}(\s_{i},\s_{j})+\sum_{i}U(\s_i)
\label{1.3}
\end{split}
\end{equation}
with $\s_i$ taking values in a finite set, the interaction kernel $\Phi^\gamma_{i-j}$
has finite range $1/\gamma$, is "smooth enough" and 
$\sum_k\|\Phi^\gamma_{k}\|_\infty =O(1)$.
No symmetry of the interaction under permutation of the phases is assumed.

Project a Gibbs-state of this model to a $\gamma$-independent sublattice $S$. 
What can we say about non-Gibbsianness? Do we expect convergence of 
the region of inverse temperatures and fields for which the decimated 
model is non-Gibbs to the corresponding mean-field values as $\gamma\downarrow 0$?

To start with, we review what is known about the translation-invariant phases. 
The work of (\cite{BoZa2},\cite{Z2}) proves the existence of 
ordered phases at low-temperatures, uniformly in (sufficiently small) $\gamma$. 
It provides essential steps of  
a Kac-Pirogov-Sinai theory, and gives a good 
understanding of the structure of the phases. 
What is lacking however is the control up to the critical temperature of 
the corresponding mean-field model. So, Kac-compatibility is only partially proved 
already on the level of translation-invariant phases. 

In order to analyze badness of a configuration $\eta$ of the 
projected lattice we need to understand whether resp. for what values 
of parameters the quenched model 
\begin{equation} \label{quenched1.3}
\frac{1}{2}\sum_{i,j\in S^c}\Phi^\gamma_{i-j}(\s_{i},\s_{j})+\sum_{i\in S^c}\Bigl( U(\s_i)
+\sum_{j\in S}\Phi^\gamma_{i-j}(\s_{i},\eta_{j})\Bigr)
\end{equation}
has a phase transition. 

This should be possible when we reverse the order of choices, taking a periodic 
$\eta$ first, and looking for values of the parameters for which there is coexistence, using 
the method of (\cite{BoZa2},\cite{Z2}). A difficulty here is that the range of temperature for which 
one is able to prove coexistence will in general be non-uniform in the period. 
Such an argument would show the existence 
of bad configurations at a discrete $\gamma$-dependent set 
of parameter-values of the model.  
So, this method would provide a proof of non-Gibbsianness 
that is not exhaustive over the whole 
presumed region of non-Gibbsianness. 
To improve this and prove non-Gibbsianness for a continuum of parameter values,
 again randomizing to choose bad configurations would be necessary. 
To prove this rigorously one would need a fully developed nonsymmetric 
random Kac-Pirogov-Sinai-theory. 
  In such a generality such a theory does not (yet) exist. 
For an outline of corresponding  results in the random field Kac Ising model (with symmetric random field distribution)  
 see \cite{BoKu05}. 
 Finally, to study a.s. Gibbsianness such a theory would have to allow 
for non-independent Gibbs-quenching, too. 
All of this would be highly non-trivial. As in the short-range situation 
the steps involving randomized bad configurations are expected 
to work only in $d\geq 3$, whereas the arguments before should be valid 
in $d\geq 2$. 

\subsubsection*{Non-Gibbsian quenched measures in mean field}

As remarked before, the non-Gibbsianness of the quenched joint measures $K(n,\sigma)$ (cf. (\ref{666},\ref{RFIMJM}))
has a mean-field analogue, as first shown in \cite{Ku3}. Physically, this again
expresses the fact that an annealed description of such measures
in terms of a Morita ``grand potential'' does not work. 
The analogy goes quite far,  in that the set of bad configurations
has measure zero or full measure in the mean-field situations according to the behaviour of the analogous model 
in finite dimensions (Zero measure but 
non-empty for dilution, full measure in the phase transition regime of the 
random field model). 

Thus for these quenched measures the connection between disorder and 
non-Gibbsianness again is ``intrinsic'', whereas disordered methods 
as a tool may or may not occur in the mean-field analysis of systems 
without disorder (in fact it occurs for the Glauber-evolved 
situation \cite{KL}, but not for decimation \cite{Ku3}).

Furthermore, for mean-field quenched models 
one can perform a rather complete analysis of the Morita
approach and where and how it breaks down and what one still can learn from it
(see \cite{Kue2,Ku4} and references mentioned there). 

Again, it would be of interest to see if one can connect the analyses 
of the mean-field and the lattice models via a Kac limit, although again 
we expect that some of the same complicated 
technical problems we mentioned before 
would need to be solved (in particular developing a general 
non-symmetric random Pirogov-Sinai theory).

\section{Summary and Conclusions}

We have reviewed two different connections which exist between the study of 
non-Gibbsian measures and of disordered systems. It turns out that there are
two main types of connections: 

\noindent
1) Results and insights developed in the study of 
disordered systems often provide useful tools to study non-Gibbsianness. 

\noindent
2) Disordered systems give rise to non-Gibbsian measures.

We have discussed these in some detail, mentioning various examples 
and also the various physical meanings of these non-Gibbsian results, 
including in renormalization group theory, non-equilibrium 
questions and the Morita approach to disordered systems.
Also we mentioned how these ideas are developed in a mean field setting, and 
we have suggested how a link may be developed in the Kac limit.











\bigskip
\noindent {\em Acknowledgements}:

We thank all our colleagues with whom we have worked on or discussed
these topics for all they have taught us, and 
Joel Lebowitz and Pierluigi Contucci for their 
invitation to contribute to this volume.

\end{document}